\documentclass[a4paper,12pt]{article}
\usepackage[top=3cm,bottom=3.4cm,left=2cm,right=2cm]{geometry}

\usepackage{amsmath,amssymb,graphicx}
\usepackage{amsbsy}
\usepackage{url}
\usepackage{caption}
\renewcommand{\title}[1]{%
	\begin{center} \Large \bf #1 \end{center}%
	}
\renewcommand{\author}[2]{%
	\vspace{1ex}
	\begin{center} { #1}  \vspace{2mm}\\ %
	  {\it #2}%
	\end{center}%
	\addvspace{\baselineskip}%
	}

\begin{document}
\newpage
\setcounter{section}{0}
\setcounter{equation}{0}
\setcounter{figure}{0}
\baselineskip 5mm
\begin{flushright}
YITP-20-39 \\
April, 2020
\end{flushright}
\vspace{10mm}
\title{%
Necessity and Insufficiency of Scale Invariance \\
for solving Cosmological Constant Problem
}%
\author{%
Taichiro Kugo\footnote{E-mail: kugo@yukawa.kyoto-u.ac.jp}
}{%
Yukawa Institute for Theoretical Physics, Kyoto University, 
Kyoto 606-8502, Japan
}%
\vspace{3ex}
\abstract{%
A scenario based on the scale invariance for explaining the vanishing 
cosmological constant (CC) is discussed. I begin with a notice on 
the miraculous fact of the CC problem that the vacuum energies totally 
vanish at each step of hierarchical and successive spontaneous symmetry 
breakings. 
I then argue that the  classical scale invariance is a {\em necessary} 
condition for the  calculability of the vacuum energy.  

Next, I discuss how {\em sufficient} the scale invariance 
is for solving the CC problem.  
First in the framework of classical field theory, the scale invariance 
is shown to give a natural mechanism for realizing the miracle of 
vanishing vacuum energies at every step of spontaneous symmetry breakings. 
Then adopting Englert-Truffin-Gastmans' prescription to maintain 
the scale invariance in quantum field theory, 
I point out that the quantum scale invariance alone is not yet 
sufficient to avoid the superfine tuning of coupling 
constants for realizing vanishingly small cosmological 
constant, whereas the hierarchy problem may be solved. 
Another symmetry or a mechanism is still necessary which protects 
the flat direction of the potential against the radiative corrections.
 }
\vfill
\begin{center}
A talk presented at {\sl Corfu Summer Institute 2019 \\
"School and Workshops on Elementary Particle Physics and Gravity" (CORFU2019)}\\
31 August - 25 September 2019, Corf\`u, Greece.  
\end{center}
\def\vec{\boldsymbol}
\def\mbf{\boldsymbol}
\def\bfe{\vec{e}}
\def\Red{}
\def\Pink{}
\def\Blue{}
\def\Green{}
\def\calA{{\cal A}}
\def\calB{{\cal B}}
\def\calC{{\cal C}}
\def\calD{{\cal D}}
\def\calE{{\cal E}}
\def\calP{{\cal P}}
\def\calQ{{\cal Q}}
\def\II{I\!I}
\def\dd{d{\cdot}d}
\def\ee{e{\cdot}e}
\def\B{{\rm B}}
\def\disp{\displaystyle}
\def\tr{\mathop{\hbox{tr}}}
\def\nn{\nonumber\\}
\def\A{{\mathcal A}}
\def\B{{\rm B}}
\def\QB{Q_{\rm B}}
\def\half{{1\over2}}
\def\VEV#1{\left\langle{#1}\right\rangle}
\def\slash#1{{\ooalign{\hfil/\hfil\crcr$#1$}}}
\newcommand{\kslash}{\slash{k}}
\newcommand{\pslash}{\slash{p}}
\newcommand{\bra}[1]{\left\langle#1\right\vert}
\newcommand{\ket}[1]{\left\vert#1\right\rangle}
\newcommand{\vs}{\vspace*{5mm}}
\newcommand{\vsp}{\vspace*{1cm}}
\def\eV{\hspace{.2pt}e\hspace{-1pt}V}
\def\MeV{Me\hspace{-1pt}V}

\newpage
\section{Introduction}

Cosmological constant problem is a dark cloud hanging over the two 
well-established theories
\begin{eqnarray}
\hbox{Quantum Field Theory} \quad  \Longleftrightarrow \quad 
\hbox{Einstein Gravity Theory}.  
\nonumber 
\end{eqnarray}
I first explain my viewpoint on {\em what is actually the problem}. 

Presently observed \Red{Dark Energy} $\Lambda_0$ looks like a 
small Cosmological Constant (CC): 
\begin{equation}
\hbox{Present observed CC}: \quad 10^{-29} {\rm gr/cm}^3 
\sim10^{-47} {\rm G\eV}^4
\sim(\Red{1\,{\rm m\eV}})^4\equiv\Lambda_0\,.
\end{equation}
I do not try to explain this tiny CC now, since it will eventually be 
explained after our CC problem is solved. However, we use it as the 
\Red{scale unit $\Lambda_0$} of our discussion in this Introduction. 

Now, from my viewpoint, 
the essential point of the CC problem is the following miraculous fact; 
that is, 
there are several dynamical symmetry breakings in this world and 
they are all accompanied by vacuum condensation energies, ranging 
over wide and hierarchical scales. Nevertheless, those vacuum 
condensation energies are almost completely canceled at each stage of 
those spontaneous symmetry breakings. 

From the success of {the Standard Model}, in particular, 
we are confident of the existence of {\em at least two} symmetry breakings:
\begin{eqnarray}
\hbox{Higgs Condensation} &:& \ -V_{\rm Higgs}\ \sim\ (\ 200\,{\rm G\eV}\ )^4\ 
\sim10^9 {\rm G\eV}^4 \sim10^{56} \Lambda_0\,,\nn
\hbox{QCD Chiral Condensation} \VEV{\bar qq}^{4/3} &:&\ -V_{\rm QCD}\ \sim\  
(\ 200\,{\rm \MeV}\ )^4 
\sim10^{-3} {\rm G\eV}^4 \sim10^{44} \Lambda_0\,. \nonumber
\end{eqnarray}
These are $10^{56}$ and $10^{44}$ times larger, respectively, 
than the present CC value $\Lambda_0$.  
Nevertheless, the fact that our calm universe exists means that 
these surely existing vacuum energies are not contributing to the 
CC at all! That is, Einstein gravity does not feel 
these condensation energies at all. If these condensation energies are 
canceled by an initially prepared ``bare cosmological constant" $c$, then, 
even these two spontaneous breakings alone imply that the cancellation must 
occur exactly over more than 56 digits. If we rephrase this fact more vividly, 
then, the Higgs condensation energy $V_{\text{Higgs}}$ and chiral condensation 
energy $V_{\text{QCD}}$ are, respectively, canceled by the bare CC 
value $c$ exactly by 12 digit and 44 digit of concrete numbers, 
respectively, as shown as follows:
\begin{eqnarray}
\begin{array}{rl}
 c (\hbox{initially prepared CC})\ \hfill & \\
  =\underbrace{654321,098765}_{12\,{\rm digits}}4321,0987654321,0987654321,0987654321,0987654321\times\Lambda_0 
 &\sim10^{56}\Lambda_0
 \\[1ex]
  c + V_{\rm Higgs}= \hspace{3.8em}
 \underbrace{4321,0987654321,0987654321,0987654321,0987654321\,}_{44\,{\rm digits}}\times\Lambda_0 &\sim10^{44}\Lambda_0
  \\[1ex]
  c + V_{\rm Higgs} +V_{\rm QCD} = \hspace{14.6em}\hbox{present Dark Energy:\ \ } 
 1\times\Lambda_0 
 &\sim\Lambda_0\ 
\end{array}
\nonumber
\end{eqnarray}
Note that the vacuum energy is almost totally canceled \Red{\em 
at each stage of spontaneous symmetry breaking} as far as in the order of 
the relevant energy scale.

In this talk, I would like to propose the {\em classical scale invariance} 
as an essential ingredient for solving the CC problem. Here 
the classical scale invariance means that the theory has {\em no 
dimensionful parameters} at all.   
Indeed in Section 2, I give an argument that the classical scale invariance 
is a {\em necessary condition} for the calculability of the vacuum energy. 
Otherwise the theory must have a bare cosmological constant term 
as a free parameter UV counterterm, implying that there is 
no hope to determine its renormalized value by calculation in the theory. 

In later Sections 3 to 5, we will discuss how the scale invariance is 
sufficient to solve the CC problem. In subsection 3.1, we discuss the 
problem in a classical field theory framework, 
namely at tree level in quantum field theory. 
There I present a scale invariant model possessing a suitable potentials, 
and show that the scale invariance gives a natural mechanism for realizing 
the miracle of vanishing vacuum energies at every step of the 
successive spontaneous symmetry breakings. 
Then, moving to quantum theory in subsections 3.2 and 3.3, I explain 
the Englert-Truffin-Gastmans'\cite{Englert1976} prescription to 
maintain the scale invariance in quantum field theory. 
In Section 4, we discuss the quantum scale invariant renormalization 
explicitly for Schaposhnikov-Zenhausern's model\cite{ShapoZen} of two 
scalar fields, Higgs and dilaton fields, whose characteristic energy 
scales are $10^2$G\eV and $10^{18}$G\eV, respectively.  Based on the explicit 
computations by Ghilencea\cite{Ghilencea1}, 
we will see that the hierarchy is maintained stable against the radiative 
corrections. However, I will point out that 
we actually need superfine tuning of coupling constants to realize 
the vanishingly small vacuum energy, implying reappearance of CC problem. 
The puzzle why the quantum scale invariance does not automatically 
guarantees the vanishing vacuum energy is resolved in Section 5. 
In Section 6, we discuss in more detail and generally how the hierarchy 
problem is solved in quantum scale invariant theory even including gravity 
loop corrections. Section 7 is devoted to conclusion.

\section{Scale Invariance is a Necessary Condition}

We show in this section that the classical scale invariance is a 
necessary condition for the CC problem to be solvable. For preparation for it, 
we first have to clear up a possible confusion about the vacuum energy.

\subsection{quantum vacuum energy $\simeq $ potential energy}

People may suspect that there are two distinct sources for the cosmological 
constant. One is the vacuum energy in quantum field theory, zero-point 
oscillation energy for boson fields and negative energy in the Dirac sea for 
fermions, 
\begin{equation}
\fbox{({Quantum}) \ Vacuum Energy}\qquad  
\sum_{{\mbf k},s} \half \hbar \omega_{\mbf k}
- \sum_{{\mbf k}, s} \hbar E_{\mbf k} 
\label{eq:QuantumVacuumEnergy}
\end{equation}
which is divergent in nature and usually simply discarded. Another is the 
potential in classical field theory: 
\begin{equation}
\fbox{({Classical}) \ Potential Energy} \qquad V(\phi_c) : \hbox{potential}
\end{equation}
which is finite in nature and gives the vacuum condensation energy in the case 
of spontaneous symmetry breaking. 
These two are separately stored in our (or my, at least) memory, 
but actually, almost the same object, as we now see. 

\def\r{{\rm r}}

We now show for the vacuum energies in the Standard Model (SM) that 
\begin{equation}
\hbox{quantum Vacuum Energy} \simeq  \hbox{Higgs Potential Energy}.
\end{equation}
To see this more explicitly, let us consider 
a simplified (analogue of) SM:
\begin{eqnarray}
{\cal L}_\r &=& \bar\psi\bigl(i\gamma^\mu\partial_\mu- y \phi(x)\bigr)\psi(x) \nn
&&+ \frac12 \bigl(\partial^\mu\phi(x)\partial_\mu\phi(x) -m^2\phi^2(x)\bigr) - \frac{\lambda}{4!}\phi^4(x)
-h m^4.
\nonumber
\end{eqnarray}
Here, $\phi$ is a single component scalar field as an analogue of Higgs field,  
and $\psi$ is a Dirac fermion as an analogue of quark/lepton fields whose mass 
comes solely from the non-vanishing vacuum expectation value (VEV) of Higgs, 
$\VEV{\phi}\not=0$. The last term $-hm^4$ is the vacuum energy (CC) term. 

Effective action and effective potential are calculated prior to the 
vacuum choice (i.e., calculable independently of the choice of the vacuum). 
The effective potential $V(\phi)$ at 1-loop level in this simplified SM 
is given in the following well-known form :
\def\MSbar{$\overline{\text{MS}}$}
\begin{align}
V(\phi,m^2) &= V_{\text{tree}} +V_{\text{1-loop}} 
+ \delta V^{(1)}_{\text{counterterms}} \nn 
V_{\text{tree}} &= 
\frac12 m^2\phi^2 + \frac{\lambda}{4!}\phi^4 + hm^4 \nn 
V_{\text{1-loop}}
&= 
\frac12 \int{d^4k\over i(2\pi)^4} \ln ( -k^2 + 
\underbrace{m^2+\half\lambda\phi^2}_{=M_\phi^2(\phi)} )
-2 \int{d^4p\over i(2\pi)^4}\ln (-p^2+\underbrace{\ y^2\phi^2\ }_{=M_\psi^2(\phi)})\,.  
\label{eq:1-loop}
\end{align}
The 1-loop integral is evaluated in the 
dimensional regularization.
Using dimensional formula
\begin{equation}
\half \,\mu^{4-n}\int{d^nk\over i(2\pi)^n} \ln ( -k^2 +M^2 )
={M^4\over64\pi^2}\Bigl(\quad -\frac1{\bar\varepsilon}\hspace{-2em}\underbrace{+\ln{M^2\over\mu^2}-\frac32}
_{\hbox{Coleman-Weinberg potential}} \Bigr), 
\end{equation}
and dropping the $1/\bar\varepsilon$ parts in \MSbar\ renormalization scheme
$\Bigl(\displaystyle\frac1{\bar\varepsilon}= \frac1{\varepsilon}-\gamma+\ln4\pi$, 
$\displaystyle\varepsilon=2-{n\over2}\Bigr)$, 
we obtain {\em finite} well-known renormalized 1-loop effective potential:
\begin{eqnarray}
V(\phi,m^2) &=&
\frac12 m^2\phi^2 + \frac{\lambda}{4!}\phi^4 + hm^4 \nn[1.5ex]
&&+ 
{(m^2+\half\lambda\phi^2)^2\over64\pi^2}\left(\ln{m^2+\half\lambda\phi^2\over\mu^2}-\frac32 \right)
-4{(y\phi)^4\over64\pi^2}\left(\ln{y^2\phi^2\over\mu^2}-\frac32 \right) 
\end{eqnarray}
Note that the divergences $\propto1/\bar\varepsilon$ appear in the terms proportional to
\begin{eqnarray}
M_\phi^4(\phi)&=& \bigl( m^2+\frac{\lambda}2\phi^2\bigr)^2 = m^4 + \lambda m^2\phi^2 
+ \frac{\lambda^2}4 \phi^4 \quad \hbox{and to} \nn
M_\psi^4(\phi)&=& \bigl( y\phi\bigr)^4 = y^4\,\phi^4\,.
\end{eqnarray} 
These divergences proportional to $\phi^4$, $m^2\phi^2$ and $m^4$ are 
renormalized into $\lambda$, $m^2$ and $h$, respectively.
Here we should recall the fact that these 1-loop contributions of the 
boson and fermion loops are just the same object as the quantum vacuum 
energies mentioned above in Eq.~(\ref{eq:QuantumVacuumEnergy}), namely,  
zero-point oscillation energy for boson fields and 
negative energy in the Dirac sea for fermions. They are divergent but are 
renormalized into the parameters $\lambda$, $m^2$ and $h$. 
The main part of quantum vacuum energies are already included in 
the classical potential $V_{\text{tree}}(\phi)$ with renormalized parameters 
$\lambda$, $m^2$ and $h$, since the 1-loop parts (i.e., 
Coleman-Weinberg potential parts) are small corrections to the renormalized 
tree level potential at energy scale around the renormalization point 
$\phi\sim\mu$.


\subsection{conclusions from these simple observations}


From this simple observation, we can 
draw very interesting and important conclusions. 
As far as the matter fields and gauge fields are concerned in the SM, 
we note that their masses solely come from the Higgs condensation $\VEV{\phi}$, 
so  
\begin{eqnarray}
\begin{minipage}[c]{.8\textwidth}  
the quantum vacuum energies coming from the matter and gauge fields 
are {\em calculable} and {\em finite} 
quantities in terms of the renormalized $\lambda$ parameters.
\end{minipage}
\nonumber 
\end{eqnarray}
This is because their masses $M$ are proportional to Higgs VEV $\phi$, and 
the divergences of their vacuum energies are proportional 
to $\phi^4$ (at 1-loop, at least.) 

However, the {\em Higgs field itself is an exception}\,!\ \  
The divergences of the Higgs vacuum energy are not only $m^2\phi^2$ 
and $\phi^4$ but also the zero-point function proportional to $m^4$. 
This comes from the right diagram in Figure 1. The left diagram proportional 
to $m^2$ vanishes as far as we use dimensional regularization.  
\captionsetup{format=hang,margin=17mm,font=footnotesize}
\begin{figure}[htb]
\begin{center}
\includegraphics[width=8cm
               ]{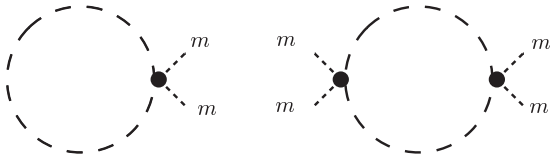}
 \caption{Divergent vacuum energy diagrams coming from the Higgs loop, 
 where the dotted line represents the massless Higgs propagators.}
\end{center}
\label{fig:1}
\end{figure}
In order to cancel 
that part, we have to prepare the bare 
vacuum energy (CC) term $h_0m_0^4$ from the beginning to yield the 
counterterm:
\begin{eqnarray}
h_0m_0^4 &=& Z_hZ_m^2\,hm^4=(1+F) hm^4 \quad \rightarrow\quad 
F^{(1)}h = {1\over64\pi^2}{1\over\bar\varepsilon}.\nonumber
\end{eqnarray}

Then, the renormalized CC term $hm^4$ becomes a {\em free parameter}.
This implies that {\em there is no chance to explain the value of CC}.

We thus reach an important conclusion: 

\begin{center}
\framebox{$
\begin{array}{l}
\hbox{For the calculability of CC, \ we should have \ \ $m^2=0$, \ \ or equivalently} \\
\hbox{no dimensionful parameters in the theory\ \ $\Rightarrow$\ \ 
{\em (Classical) Scale-Invariance}.}
\end{array}
$}
\end{center}

%

\section{Scale Invariance may solve the CC Problem}

Our world is almost scale invariant: 
that is, the SM Lagrangian is 
scale invariant {\em except for the Higgs mass term}. 
So if the Higgs mass term comes from the spontaneous breaking of scale 
invariance at higher energy scale physics, the total system can 
really be scale invariant (classically, at least):
\begin{equation}
\lambda(h^\dagger h-m^2)^2 \quad \rightarrow\quad \lambda(h^\dagger h-\varepsilon\Phi^2)^2.
\end{equation}
where $h$ is Higgs field and $\Phi$ is a certain scalar field relevant to the 
higher energy physics; for instance, $\Phi$ may be a field appearing in front 
of the Einstein-Hilbert term as 
\begin{equation}
\int d^4x \sqrt{-g} \  \Phi^2\,R\,. 
\end{equation}
We call this field $\Phi$ dilaton henceforth since it becomes the Nambu-Goldstone 
boson for spontaneous breaking of scale ($=$ dilatation) invariance by 
its non-vanishing VEV. 

I will explain in this section how the classical scale invariance (SI) 
may solve 
the CC problem; in particular, it would give a natural mechanism why the 
vacuum energy remains vanishing at every stage of hierarchical successive 
spontaneous symmetry breakings. 
Similar ideas have been proposed so far by many authors including 
Shaposhnikov and Zenhausern\cite{ShapoZen}, 
Antoniadis and Tsamis\cite{Antoniadis}, 
Tomboulis\cite{Tomboulis},
Wetterich\cite{Wetterich}, and 
others\cite{Shaposhnikov:2008xb,Rabinovici:1987tf,Meissner:2006zh,%
Ferreira:2016vsc,Oda:2018lgm}. 
My scenario is most similar to 
Shaposhnikov and Zenhausern\cite{ShapoZen}, but no one has ever pointed 
out that it gives a natural mechanism for realizing vanishing vacuum energy 
at every stage of successive spontaneous symmetry breakings. 

Before explaining my scenario using {\em global SI}, let me mention to the 
work by 
Antoniadis and Tsamis\cite{Antoniadis} and Tomboulis\cite{Tomboulis}, 
whose papers appear very early and actually 
contain almost all basic ideas in this direction for solving the CC problem. 
Nevertheless those work use {\em local SI} which I think has to be useless: 
\begin{eqnarray}
\fbox{Local SI theory with dilaton (without Weyl gauge field) 
is meaningless.}
\nonumber
\end{eqnarray}

The reason is the following. 
If the dilaton field $\Phi_0(x)$ with dimension one is present, 
{\em any action} can be cast into local SI form so that the local SI itself 
means nothing: Indeed, for any given 
action $S[\phi]$ which may contain any mass terms, we can replace 
any fields $\phi_i$ with dimension $d_i$ by the scale invariant 
fields 
\begin{equation}
\phi_i \ \rightarrow\ \Phi_0^{-d_i}\phi_i =: \phi'_i.
\end{equation}  
Then the action becomes local scale invariant under $\phi_i(x)\rightarrow\lambda 
(x)^{d_i}\phi_i(x)$ and $\Phi_0(x)\rightarrow\lambda(x)\Phi_0(x)$. But this local symmetry is
{\em fake} since the system reduces to the original action $S[\phi]$ in 
the unitary gauge $\Phi_0(x)=1$. q.e.d.

Essentially the same but more detailed discussion was given by 
Tsamis and Woodard\cite{TsamisWoodard} for the conformal scalar-metric 
gravity theory.
Note, however, that this argument applies only to the local SI system in which 
Weyl gauge field is absent. If the Weyl gauge field exists in the system, 
the gauge fixing cannot eliminate all four components of the Weyl gauge field
\cite{GhilenceaWeyl, OdaWeyl}. 

So, by SI henceforth, we always mean {\em global scale invariance}, 
or equivalently, the 
{absence of dimensionful parameters} in this paper.

\subsection{classical scale invariance: a possible scenario}

Suppose that our world has \Red{no dimensionful parameters}. Let 
the effective potential $V$ of the total system look like
\begin{eqnarray}
\begin{array}{cccccc}
V(\phi) = & V_0(\Phi) & + &V_1(\Phi, h)& + &V_2(\Phi, h, \varphi) \\  
    & \downarrow &   & \downarrow       &   &   \downarrow      \\  
    & M      & \gg&  \mu     & \gg&   m           \\  
\end{array}
\nonumber
\end{eqnarray}
We suppose that $V_0(\Phi), V_1(\Phi, h)$ and $V_2(\Phi, h, \varphi)$ are relevant to
the physics at three energy scales, Planck scale $M$, electroweak scale $\mu$ and 
QCD scale $m$, respectively, although they contain no dimensionful parameters.  
We also suppose that $h$ and $\varphi $ are Higgs field and chiral $SU(2)$ 
sigma-model 
field, respectively.
Then, \Red{classically}, it satisfies the scale invariance relation :
\begin{equation}
\sum_i \phi^i{\partial\over\partial\phi^i}\  V(\phi) = 4V(\phi),
\label{eq:CS}
\end{equation}
with $\phi_i$ representing all the relevant scalar fields collectively. This 
implies that the vacuum energy vanishes at any stationary point 
$\VEV{\phi_i}=\phi^0_i$:
\begin{eqnarray}
V(\phi^0)=0.
\nonumber
\end{eqnarray}
Important point is that {\em this holds at every stage of spontaneous 
symmetry breakings} as far as the potential $V_0(\Phi)$, \  
$V_0(\Phi)+V_1(\Phi, h)$ and 
$V_0(\Phi)+V_1(\Phi, h)+V_2(\Phi, h, \varphi)$ are separately scale invariant 
(i.e., of dimension 4). 
This should be so because, for instance, we can retain only 
$V_0(\Phi)$ part when discussing 
the physics at scale $M$ since
$h$ and $\varphi$ are expected to get much smaller VEVs of order $\mu$ 
or lower. 
Then the scale invariance guarantees $V_0(\Phi_0)=0$. 
This thus gives a very natural mechanism for realizing the {\em miracle} that 
the vacuum energy remains vanishing at every step of 
spontaneous symmetry breakings. 

We can now write a toy model of potentials. First part is 
\begin{eqnarray}
\fbox{$V_0(\Phi)= \half\lambda_0 ( \Phi^2_1 - \varepsilon_0\Phi_0^2)^2,$}
\nonumber
\end{eqnarray}
in terms of two real scalar fields $\Phi_0$ and $\Phi_1$, to realize VEVs
\begin{equation}
\VEV{\Phi_0}=M \quad \hbox{and}\quad 
\VEV{\Phi_1}=\sqrt{\varepsilon_0} M \equiv M_1. 
\end{equation}
This $M$ is totally spontaneous and there is no meaning in its magnitude 
at this stage. Only meaningful is whether it vanishes 
or not. We suppose $M$ be \Red{Planck mass} giving the Newton 
coupling constant via the scale invariant Einstein-Hilbert term
\begin{eqnarray}
S_{\rm eff} &=& \int d^4x \, \sqrt{-g} \Bigl\{ 
c_1 \Phi_0^2\, R + 
c_2 R^2 + c_3 R_{\mu\nu}R^{\mu\nu} + \cdots\Bigr\} \,.
\nonumber 
\end{eqnarray}
If the grand unified theory (GUT) stage exists, $\varepsilon_0$ may be a constant 
as small as $10^{-4}$ 
and then $\Phi_1$ gives the scalar field which breaks 
GUT symmetry; 
e.g., $\Phi_1:{\bf 24}$ causing $SU(5)\rightarrow SU(3)\times SU(2)\times U(1)$. 

$V_1(\Phi, h)$ part causes the electroweak symmetry breaking:
\begin{eqnarray}
\fbox{$V_1(\Phi, h)= \half\lambda_1\left( h^\dagger h - \varepsilon_1\Phi_1^2 \right)^2$}
\nonumber
\end{eqnarray}
with very small parameter $\varepsilon_1\simeq(10^2\text{G\eV}/10^{16}\text{G\eV})^2
\simeq10^{-28}$. 
This reproduces the Higgs potential when $h$ is 
the Higgs doublet field and $\varepsilon_1\Phi_1^2$ term is replaced by the VEV 
$\varepsilon_1M_1^2 =\mu^2/\lambda_1\sim(10^{2}\text{G\eV})^2$. 

$V_2(\Phi, h, \varphi)$ part 
causes the chiral symmetry 
breaking, e.g., SU(2)${}_{\rm L}\times$SU(2)${}_{\rm R}\ \rightarrow\ $ SU(2)${}_{\rm V}$. 
Using the $2\times2$ matrix scalar field 
$\varphi=\sigma+i{\mbf\tau}\cdot{\mbf\pi}$ (chiral sigma-model field), 
we may similarly write 
the potential 
\begin{eqnarray}
\fbox{$V_2(\Phi, h, \varphi)= \frac14 \lambda_2\left( \tr(\varphi^\dagger\varphi ) 
-\varepsilon_2\Phi_1^2 \right)^2 + 
V_{\rm break}(\Phi, h, \varphi)$}
\nonumber
\end{eqnarray}
with another small parameter $\varepsilon_2\simeq 10^{-34}$. 
The first term reproduces the linear $\sigma$-model potential invariant 
under the chiral SU(2)${}_{\rm L}\times$SU(2)${}_{\rm R}$ transformation 
$\varphi\rightarrow g_{\rm L}\varphi g_{\rm R}$ when 
$\varepsilon_2\Phi_1^2$ is replaced by the VEV 
$\varepsilon_2M_1^2 =m^2/\lambda_2$. 
The last term $V_{\rm break}$ stands for the chiral symmetry breaking 
term which is caused by the explicit quark mass terms appearing as the result 
of tiny Yukawa couplings of $u, d$ quarks, $y_u, y_d$, 
to the Higgs doublet $h$; e.g.,
\begin{eqnarray}
V_{\rm break}(\Phi, h, \varphi)= \half \varepsilon_3\Phi_1^2 
\tr\left[
\varphi^\dagger 
\begin{pmatrix} y_u \tilde h\, & \,y_d h \\
\end{pmatrix} + \hbox{h.c.}  
\right]
\nonumber
\end{eqnarray} 
with $\varepsilon_3\sim4\pi\varepsilon_2$ and $\tilde h\equiv i\sigma_2h^*$. 

\subsection{quantum mechanically}

As soon as we come to quantum field theory, we are confronted 
with the SI anomaly: 
\begin{eqnarray}
\fbox{
Scale invariance suffers from an Anomaly.
}
\nonumber 
\end{eqnarray}
Usual wisdom tells us so. Owing to the UV divergence in quantum field theory, 
it is necessary to introduce a (dimensionful) renormalization point $\mu$, 
which necessarily break the classical SI. 
If we also take the renormalization point 
$\mu$ into account, the dimension counting identity comes to read
\begin{eqnarray}
\left(\mu{\partial\over\partial\mu}+\sum_i \phi_i{\partial\over\partial\phi_i}\right)V(\phi)=4V(\phi).
\label{eq:DimCount}
\end{eqnarray}
The anomaly $\mu(\partial/\partial\mu)V$ term may be eliminated by using 
renormalization group equation (RGE): 
\begin{eqnarray}
\left(\mu{\partial\over\partial\mu}+\sum_a\beta_a(\lambda){\partial\over\partial\lambda_a}+\sum_i\gamma_i(\lambda)\phi_i{\partial\over\partial\phi_i}
\right)V(\phi)=0.
\nonumber
\end{eqnarray}
Then, we obtain 
\begin{eqnarray}
\left(\sum_i (1-\gamma_i(\lambda))\phi_i{\partial\over\partial\phi_i}-\sum_a\beta_a(\lambda){\partial\over\partial\lambda_a}\right)
V(\phi)=4V(\phi),
\label{eq:AnomDimCount}
\end{eqnarray}
which replaces the above naive dimension counting equation 
$\sum_i \phi_i(\partial/\partial\phi_i) V(\phi)=4V(\phi)$.
With either Eq.~(\ref{eq:DimCount}) or Eq.~(\ref{eq:AnomDimCount}), we cannot 
conclude the vanishing potential value $V(\phi^0)=0$ at the stationary point 
$\phi^0$.  
Eq.~(\ref{eq:AnomDimCount}) shows that the anomalous dimension $\gamma_i(\lambda)$ 
is not the problem, but $\beta_a(\lambda)$ terms may be problematic.

Still, if we assume the existence of {\em Infrared Fixed Points} 
$\beta_a(\lambda_{\rm IR})=0$ and that the theory on top of that point $\lambda_{\rm IR}$ 
is well-defined, then, I can prove 
that the potential value $V(\phi^0)$ at the stationary point 
$\phi=\phi^0$ is zero at any finite $\mu$ (not necessarily in the IR limit 
$\mu\rightarrow0$).  
That is, {\em the vanishing property of the stationary potential value 
$V(\phi)$ is not injured} by the scale-invariance anomaly\cite{Kugo2017}.

Even then, however, we will meet the same difficulty -- 
flat direction problem -- 
as that we will encounter 
also in the next approach which we discuss from now on. So we do not 
discuss this approach assuming the IR fixed point anymore here.


\subsection{quantum scale invariant renormalization}

Shaposhnikov and Zenhausern\cite{ShapoZen} 
proposed a new approach to this anomaly 
obstacle for the scale invariance scenario.  
Their proposal is based on a very simple observation that 
{\em SI can be maintained even in quantum field theory if we have a 
dilaton field  $\Phi$} in the system. 
Generally, if a regularization method exists which keeps a symmetry, then 
it implies the absence of anomaly for the symmetry. 
In this case of SI,  
the extension to $n$-dimension is shown possible keeping SI 
if a dilaton field $\Phi$ is used, as we explain shortly. 

This way of quantum SI renormalization is, however, not new, 
but actually has long been known since the original proposal by 
Englert, Truffin and Gastmans\cite{Englert1976}. It was also used in 
the prior scale invariant 
approaches to the CC problem\cite{Antoniadis,Tomboulis}.

Recall the way how the scalar quartic coupling $\lambda$ and Yukawa coupling $y$ 
are kept dimensionless in $n$ dimension in the usual dimensional 
regularization. It is realized by introducing renormalization scale $\mu$ 
as follows:
\begin{eqnarray}
&&\hbox{\bf Usual dimensional regularization} \nn 
&&\hspace{3em}\lambda\,(h^\dagger(x)h(x))^2 \quad \ \,\rightarrow\quad \lambda\,\mu^{4-n}(h^\dagger(x)h(x))^2  \qquad [h]={n-2\over2}\nn
&&\hspace{3em}y\, \bar\psi(x)\psi(x)h(x) \quad \rightarrow\quad y\,\mu^{{4-n\over2}}\bar\psi(x)\psi(x)h(x) \qquad [\psi]={n-1\over2}  
\end{eqnarray}
To avoid the introduction of explicit dimensionful parameter $\mu$ violating 
the SI, we can replace $\mu$ by a power of 
the dynamical dilaton field $\Phi(x)$, \ $\Phi^{\frac2{n-2}}(x)$, of dimension 1 
as  
\begin{eqnarray} 
&&\hbox{{\bf SI prescription}}\ \ \nn 
&&\hspace{3em}\lambda\,(h^\dagger(x)h(x))^2 \quad \ \,\rightarrow\quad \lambda\,[\Phi(x)^{\frac2{n-2}}]^{4-n}\,(h^\dagger(x)h(x))^2  \nn
&&\hspace{3em}y\, \bar\psi(x)\psi(x)h(x) \quad \rightarrow\quad y\,
[\Phi(x)^{\frac2{n-2}}]^{\frac{4-n}2}\,\bar\psi(x)\psi(x)h(x)\,.\hspace{3.2em}  
\end{eqnarray}
Since no dimensionful parameter is introduced, this prescription 
really keeps SI in any dimension $n$. But the price we have to pay is the 
non-renormalizable interaction terms; that is,  
on the vacuum in which the dilaton field develops the VEV $\VEV{\Phi(x)}=M$, 
the introduced fractional power of $\Phi(x)$ yields 
{\em non-polynomial ``evanescent" interactions} $\propto2\epsilon=4-n$:
\begin{eqnarray}
&&\Phi(x)=M+\phi(x) \ \ \rightarrow\ \ 
[\Phi(x)]^{{4-n\over n-2}} = M^{\epsilon\over1-\epsilon}\left( 1+ 
{\epsilon\over1-\epsilon}{\phi(x)\over M}+
+\half\frac{\epsilon(2\epsilon-1)}{(1-\epsilon)^2}{\phi(x)^2\over M^2}+\cdots\right).
\nonumber
\end{eqnarray}
This prescription gives {quantum scale invariant} theory, which might  
realize the vanishing CC. So let us examine this theory in more detail. 

\section{Quantum Scale-Invariant Renormalization: 2-scalar model}

Explicit calculations were performed by Ghilencea and his 
collaborators\cite{Ghilencea1,GLO,Ghilencea2}
in a simple \Red{2-scalar model}; 
following Ref.~\cite{Ghilencea1}, we henceforth 
use notations $\phi(x)$ and $\sigma(x)$ to denote our Higgs field $h(x)$ 
and dilaton field $\Phi(x)$ ($h\ \rightarrow\ \phi,\ \Phi\ \rightarrow\ \sigma$). Then the Lagrangian 
reads
\begin{equation}
{\cal L}= \half \partial_\mu\phi\cdot\partial^\mu\phi+ 
\half \partial_\mu\sigma\cdot\partial^\mu\sigma- V(\phi,\sigma) 
\end{equation}
with scale-invariant potential in $n$ dimension:
\begin{equation}
V(\phi,\sigma) = \mu(\sigma)^{4-n}\left(\frac{\lambda_\phi}4 \phi^4 - \frac{\lambda_m}2\phi^2\sigma^2+
\frac{\lambda_\sigma}4 \sigma^4\right)
\end{equation} 
with 
\begin{equation}
\mu(\sigma)= z \sigma^{\frac2{n-2}}. 
\end{equation}
Here $z$ is a renormalization point parameter introduced by 
Tamarit\cite{Tamarit} to discuss renormalization group equation (RGE) 
in this quantum SI theory, but we can 
take $z=1$ if we do not care about RGE. 
At tree level, $\lambda_m^2=\lambda_\phi\lambda_\sigma$ is assumed so that the potential becomes 
a complete square form:
\begin{eqnarray}
&&V(\phi,\sigma) = \mu(\sigma)^{4-n}\frac{\lambda_\phi}4\left( \phi^2 - \varepsilon\sigma^2\right)^2, \nn
&&\ \ \text{with}\qquad \lambda_m = \varepsilon\lambda_\phi, \qquad \lambda_\sigma=\varepsilon^2 \lambda_\phi.
\label{treePotential}
\end{eqnarray}
Note: If the dilaton $\sigma$ and the Higgs $\phi$ are supposed 
to get the VEVs of order of 
the Planck scale mass $M\sim10^{18}$G\eV and the electroweak mass 
$\mu\sim10^2$G\eV, respectively, then the parameter $\varepsilon=\VEV{\phi}^2/\VEV{\sigma}^2$ 
is very tiny $\sim10^{-32}$. We know that Higgs quartic coupling 
$\lambda=2\lambda_\phi\sim1/4$ so that $\lambda_m$ and $\lambda_\sigma$ are very tiny of 
$O(\varepsilon)$ and $O(\varepsilon^2)$, respectively.  

Ghilencea has shown the following for this quantum scale invariant theory:
\begin{enumerate}
\item {\em Non-renormalizability}: higher and higher order non-polynomial 
interaction terms of the form
\begin{equation}
\frac{\phi^{4+2p}}{\sigma^{2p}} \quad ( p=1,2,3,\cdots)
\end{equation} 
are induced by the {\em evanescent interactions} at higher loop level 
(up to $p\leq\ell$ at $\ell$ loop level),  
and they must also be included as counterterms. 
These terms, however, can be neglected in the low-energy region 
below Planck scale $E< \VEV{\sigma}\sim M$. So the usual renormalizable theory 
is an effective low energy theory valid below Planck energy, irrespectively 
of whether the gravity is quantized or not.

\item {\em Mass hierarchy is stable}: If we put 
\begin{equation}
\lambda_\phi= \bar{\lambda}_\phi, \quad \lambda_m=\varepsilon\bar{\lambda}_m, \quad \lambda_\sigma=\varepsilon^2\bar{\lambda}_\sigma 
\label{eq:CouplingOrder}
\end{equation}
with $O(1)$ coupling constants $\bar{\lambda}_i$ $(i=\phi,m,\sigma)$ and very tiny 
$\varepsilon= 10^{-32}$,
then, $\bar{\lambda}_i$'s remain of $O(1)$ stably against radiative corrections.    
This is essentially because $\sigma^2\phi^2$ term comes only through the 
$\lambda_m\phi^2\sigma^2$ interaction. 
\end{enumerate}

Explicit form of the one-loop potential at $n=4$ is actually given 
in the {\em Scale Invariant} form:
\begin{eqnarray}
V(\phi,\ \sigma) &=& \frac{\lambda_\phi}4 \phi^4 - \frac{\lambda_m}2\phi^2\sigma^2+
\frac{\lambda_\sigma}4 \sigma^4 \nn
&&{}+\frac{\hbar}{64\pi^2}\left\{
M_1^4\biggl(\ln\frac{M_1^2}{z^2\sigma^2}-\frac32\biggr)
+M_2^4\biggl(\ln\frac{M_2^2}{z^2\sigma^2}-\frac32\biggr)
+\Delta V
\right\}, 
\label{eq:1-loopPot} \\[1ex]
&&\Delta V = -\lambda_\phi\lambda_m\frac{\phi^6}{\sigma^2}+(16\lambda_\phi\lambda_m-6\lambda_m^2+3\lambda_\phi\lambda_\sigma)\phi^4 \nn
&& \hspace{3em}{}+( -16\lambda_m+25\lambda_\sigma)\lambda_m \phi^2\sigma^2-21\lambda_\sigma^2\sigma^4.
\end{eqnarray}
where $M_i^2$ $(i=1,2)$ are two mass-square eigenvalues for two scalar fields 
around the VEVs $\phi$ and $\sigma$, so $M_i^2/\sigma^2$ are dimensionless functions of 
dimensionless variable $\phi^2/\sigma^2$. The $\Delta V$ potential is the finite part 
which comes from the $O(\varepsilon)$ evanescent interaction terms multiplied by 
the one-loop divergence $1/\varepsilon$. 

However, there is a problem to which Ghilencea has not mentioned:
\begin{enumerate}
%
%
\item[3.] {\em Vanishing CC again requires fine tuning}! owing to quantum 
corrections.
\end{enumerate}
This is the most important point in this paper, so let us now explain it 
in detail. 
Since $V(\phi, \sigma)$ is a dimension-4 function in $\sigma$ and $\phi$, it 
takes the form 
\begin{equation}
V(\phi, \sigma)= \sigma^4 W(x)\quad  \hbox{with}\quad  x\equiv\phi^2/\sigma^2. 
\end{equation}
Since the stationarity conditions
\begin{eqnarray}
\left\{
\begin{array}{rcccll}
\displaystyle \phi{\partial\over\partial\phi}V&=& \displaystyle \sigma^4 W'(x) \cdot 2x &=0   \\[2ex]
\displaystyle \sigma{\partial\over\partial\sigma}V&=& \sigma^4\Bigl( 4W(x) +W'(x)\cdot(-2x)\Bigr)&=0
\end{array}\right.
\label{eq:Weql0}
\end{eqnarray}
requires both
\begin{equation}
W'(x)=0\quad \hbox{and}\quad  W(x)=0 \quad \hbox{are satisfied, \ \ unless}\ \ \sigma=\phi=0.
\end{equation} 
Let us examine these conditions with the above 1-loop potential (\ref{eq:1-loopPot}):
\begin{eqnarray}
W(x) 
 &=& \frac{\lambda_\phi}4 x^2 - \frac{\lambda_m}2 x + \frac{\lambda_\sigma}4  \nn
&&{}+\frac{\hbar}{64\pi^2}\biggl\{
\frac{M_1^4}{\sigma^4}\biggl(\ln\frac{M_1^2}{z^2\sigma^2}-\frac32\biggr)
+\frac{M_2^4}{\sigma^4}\biggl(\ln\frac{M_2^2}{z^2\sigma^2}-\frac32\biggr) \nn
&& \hspace{2em}{} 
-\lambda_\phi\lambda_m x^3+(16\lambda_\phi\lambda_m-6\lambda_m^2+3\lambda_\phi\lambda_\sigma)x^2  
+( -16\lambda_m+25\lambda_\sigma)\lambda_m x - 21\lambda_\sigma^2
\biggr\}\,. \nonumber
\end{eqnarray}

First consider these conditions at tree level; the stationary point 
$x=x_0$ should satisfy
\begin{eqnarray}
\left\{
\begin{array}{rcl}
W'(x_0)=\displaystyle \frac{\lambda_\phi}2 x_0-\frac{\lambda_m}2=0 \ \ &\rightarrow&\ \  
\displaystyle x_0=\frac{\lambda_m}{\lambda_\phi}\ , \\[2ex]
W(x_0)=\displaystyle 
\frac{\lambda_\phi}4x_0^2-\frac{\lambda_m}2x_0+\frac{\lambda_\sigma}4=0 \ \ &\rightarrow&\ \  
\displaystyle \lambda_\sigma=\frac{\lambda^2_m}{\lambda_\phi}\ . 
\end{array}\right.
\end{eqnarray}
Note here that the stationary point $x_0=\frac{\VEV{\phi}^2}{\VEV{\sigma}^2}$ is 
already determined by the first condition $W'(x_0)=0$ alone, 
while the second one \Red{$W(x)=0$ imposes a {\em constraint on the coupling 
constants} $\lambda_i$'s}. This constraint $\lambda_\sigma\lambda_\phi=\lambda_m^2$ at this stage is the 
condition we have initially imposed on the tree potential 
in Eq.~(\ref{treePotential}).

At one-loop level, next, the stationary point may be shifted and 
the coupling constants may be adjusted:
\begin{equation}
x=x_0+\hbar x_1, \qquad \lambda_i \  \Rightarrow\ \lambda_i +\hbar\, \delta\lambda_i \ \ (i=\phi,m,\sigma)\ .
\end{equation}  
The first condition $W'(x)=0$ requires, for $O(\hbar)$ parts,
\begin{eqnarray}
W'(x)\Bigr|_{O(\hbar)}&=& 
\frac{\lambda_\phi}2 x_1 
+ \frac{\delta\lambda_\phi}2x_0 + \frac{\delta\lambda_m}2 
\nn 
&&+\frac1{64\pi^2}
\biggl[4\lambda_\phi\lambda_m(3+2x_0-x_0^2) \biggl(\ln\frac{2\lambda_m(1+x_0)}{z^2}-1\biggr)
+16\lambda_m^2(1+x_0)\biggr]\ .
\label{eq:WprimeAt1-loop}
\end{eqnarray}
This determines, as at tree level, the VEV's shift at 1-loop level $\hbar x_1$. 
We may or may not adjust the coupling constants at this stage. But important 
is the point that this condition (\ref{eq:WprimeAt1-loop}) is {\em consistent}
 with the VEV (mass) hierarchy; that is, no fine tuning of the coupling 
constants is necessary to maintain the tiny ratio of field VEVs   
$x={\VEV{\phi}^2}/{\VEV{\sigma}^2}$ at tree level, $x_0=\lambda_m/\lambda_\phi=\varepsilon\sim10^{-32}$. 
Indeed, every term in Eq.~(\ref{eq:WprimeAt1-loop}) is $O(\varepsilon)$ simply by 
keeping the order of magnitude of the coupling constants as in 
Eq.~(\ref{eq:CouplingOrder}); namely, the barred coupling constants 
are all kept of order 1, $\bar\lambda_i\sim\delta\bar\lambda_i\sim O(1)$, and then the VEV 
ratio $x_0+\hbar x_1$ consistently remains of $O(\varepsilon)$:     
\begin{equation}
\lambda_m, \delta\lambda_m\ \sim\ O(\varepsilon),\ \ \lambda_\phi, \delta\lambda_\phi\ \sim\ O(1)\ \ 
\rightarrow\ \ x_0,\ x_1 \ \sim\ O(\varepsilon)\ \ 
\rightarrow\ \ x=\frac{\VEV{\phi}^2}{\VEV{\sigma}^2} = x_0+\hbar x_1 \ \sim\ O(\varepsilon).
\end{equation}

Next is the second condition: 
\begin{eqnarray}
W(x)\Bigr|_{O(\hbar)}&=&   
 \frac{\delta\lambda_\phi}4x^2_0 + \frac{\delta\lambda_m}2 x_0+ \frac{\delta\lambda_\sigma}4 
+\frac1{64\pi^2}
\biggl[4\lambda_m^2(1+x_0)^2 
\biggl(\ln\frac{2\lambda_m(1+x_0)}{z^2}-\frac32\biggr)\biggr].
\end{eqnarray}
Again, this gives a constraint on the coupling constants. Here again 
all the terms are {\em consistently} of $O(\varepsilon^2)$, so that 
$W(x)=0$ can be realized up to $o(\varepsilon^2)$ by $O(1)$ tuning of 
the barred coupling constants 
$\bar\lambda_\phi,\ \bar\lambda_m,\ \bar\lambda_\sigma$.  
{\em However}, although $W(x)$ at the stationary point can be made vanish 
very precisely in the order as tiny as $\varepsilon^2\sim10^{-64}$, 
the {\em Vacuum Energy} $V=\sigma^4W(x)$ itself vanishes 
\Red{only} in the sense of $O(\varepsilon^2)\times\sigma^4=O((100\text{G\eV})^4)$. This is 
because the Planck energy is so huge; 
$\VEV{\sigma}^4=(10^{18} \text{G\eV})^4=10^{64}\times(100\text{G\eV})^4$. 
If we require the vanishingness up to the order of presently observed 
vacuum energy $\Lambda_0 \sim(1\,\text{m\eV})^4 \sim10^{-56}\times(100\text{G\eV})^4$, then, 
we have still to tune the barred coupling constants 
$\bar\lambda_\phi,\ \bar\lambda_m,\ \bar\lambda_\sigma$ in 56 digits!  
We still need {\em superfine tuning} even in quantum scale-invariant theory.
This is nothing but reappearance of the original CC problem! 
We have to conclude: 
\begin{equation}
\fbox{
Quantum SI is not enough to solve the CC problem. 
}
\end{equation}
Shaposhnikov and Zenhausern\cite{ShapoZen} noted that there are 
degrees of freedom of coupling constants to realize the condition $W(x)=0$,
but they did not recognize that it requires the superfine tuning of 
the coupling constants which is essentially the same problem as the CC problem 
we originally wanted to solve.

Note also, however, that this is in fact the {\em problem beyond the 
perturbation theory}. 
We are discussing the vacuum energy in much finer precision 
than the perturbation (loop) expansion parameter $\hbar/16\pi^2\sim1/158$.



\section{What happens?}  

If the theory is quantum scale-invariant, then we have the dimension counting equation 
\begin{equation}
\sum_i \phi_i\frac{\partial}{\partial\phi_i}\, V(\phi) = 4 V(\phi)
\end{equation}
which implies $V(\phi_i^0)=0$ at any stationary point $\phi_i^0$, and 
any point in that direction, $\rho\phi_i^0$ with $\forall\rho\in{\bf R}$ also 
realizes the vanishing energy $V(\rho\phi_i^0)=\rho^4V(\phi_i^0)=0$ 
({\em flat direction}). 
Conversely speaking, therefore, if $V(\phi)\not=0$ at $\exists\phi$, then the potential 
is not stationary at that $\phi$. 

In the above: $V(\phi,\sigma)=\sigma^4W(x)$ was flat in the
direction $\phi^2/\sigma^2=x_0$ at tree level, $W(x_0){=}0$, but, 
at one-loop, the potential did not exactly satisfy $W(x_0+\hbar x_1)=0$ 
at the `stationary point' realizing $W'(x_0+\hbar x_1)=0$ exactly, unless 
the coupling constants were superfine-tuned. The value 
$W(x_0+\hbar x_1)$ is just as tiny as $\varepsilon^2\sim10^{-64}$ but not exactly zero. 

This means from the above 
Eq.~(\ref{eq:Weql0}) that the point $x_0+\hbar x_1$ realizes the stationarity 
with respect to $\phi$ but not necessarily to $\sigma$. ``$W(x_0+\hbar x_1)\not=$ 
exactly 0" means that the potential has a {\em tiny} gradient 
$\sigma(\partial/\partial\sigma)V= 4\sigma^4W(x)= \sigma^4O(\varepsilon^2)\not=0$ 
in the $\sigma$-direction (or, more precisely, $\phi^2/\sigma^2=x_0+\hbar x_1$ direction)
and that the potential is actually stationary {\em only at the 
origin $\sigma=0$}! That is, 
\begin{equation}
\fbox{The flat direction is {\em lifted} by the radiative correction.}
\end{equation}
Quantum scale invariance alone does not protect the flat direction, 
automatically. Artificial superfine tuning of the coupling constants 
was required to keep the flat direction. 

Tomboulis\cite{Tomboulis} however proposed an interesting mechanism 
with which 
the condition $W(x)=0$ may automatically be satisfied without any fine  
tuning of the coupling constants. Let us explain his arguments.  
Recall that the stationarity ${\delta V\over\delta\phi}=0$ and ${\delta V\over\delta\sigma}=0$ required, 
for $V=\sigma^4W$,
 respectively, 
\begin{eqnarray}
(1) \quad  W'(x; \lambda)\Big|_{x=x_0(\lambda)}=0 \qquad  &\ \Rightarrow\ &\quad (2)\quad \ W\bigl(x_0(\lambda); \lambda\bigr) =0 
\label{eq:WprmWcond}\\
\hbox{determines the VEV ratio}\ x=\frac{\phi^2}{\sigma^2}=x_0(\lambda)\qquad  
& & \qquad \hbox{demands \Red{super fine tuning of $\lambda$'s.}}\nonumber
\end{eqnarray}
Tomboulis introduced a renormalization point $\mu$ in addition to 
the dilaton field $\sigma$ and consider the running of coupling 
constant:\footnote{%
The coupling constants run even in quantum scale invariant theory despite 
the fact that the usual beta functions $\beta_a(\lambda)$ represent the anomaly 
for the scale invariance. Tomboulis as well as Shaposhnikov-Zenhausern did 
know this fact, but it was fully clarified by Tamarit\cite{Tamarit} and 
properly used by Ghilencea\cite{Ghilencea1}. 
Here we use the letter $z$ following Ghilencea\cite{Ghilencea1} 
to denote the renormalization 
point parameter $\xi$ of Tamarit's\cite{Tamarit}.}
\begin{eqnarray}
\bar \lambda(z), \qquad z \equiv\frac{\mu}{\sigma}\ : \ 
&&\hbox{renormalization point parameter}\,. 
\end{eqnarray}
Then, he claims that, the second condition in Eq.~(\ref{eq:WprmWcond}), 
now reading 
\begin{equation}
W( x_0(\bar\lambda(z)) ;  \bar\lambda(z) ) = 0,
\end{equation}
is simply an equation determining the renormalization point $z_0=\mu_0/\sigma_0$ 
and so is automatically satisfied without any fine tuning. 

This interesting idea, however, does not work unfortunately, since
\begin{equation}
\frac{d}{dz}
W( x_0(\bar\lambda(z)) ;  \bar\lambda(z) ) = 0.
\end{equation} 
Changing the renormalization point $z=\mu/\sigma$ cannot change the value 
of $W( x_0(\bar\lambda(z)) ;  \bar\lambda(z) ) $ since it is a physical quantity 
independent of the choice of renormalization point. 

We thus have no mechanism which can preserve the flat direction against 
quantum radiative corrections. 
We still need another \Red{symmetry} to realize the flat direction. 
{\em Supersymmetry} (SUSY) would be an immediate candidate for it. 
But it will also introduce another problem how to break it spontaneously. 
Depending on the way of breaking, the superfine tuning problem may reappear.

\section{Discussions: hierarchy problem}

Although the quantum scale invariance is not yet sufficient for solving 
the CC problem, 
it should be emphasized that it already almost solved the hierarchy problem. 

There are two aspects of the hierarchy problem: 
\begin{enumerate}
\item \Blue{Origin}:\ \ to explain the origin why the hierarchy exist.
\item \Blue{Stability}:\ \  to explain its stability against radiative corrections, once it exists anyway.
\end{enumerate}

\subsection{stability against radiative corrections}

As for the stability against the radiative correction, 
it is guaranteed, for instance, by SUSY as a well-known example. 
If the system has scale invariance, there exist 
only the logarithmic divergences but \Red{no quadratic divergences}, 
so that the stability is automatic in the SM, as 
was emphasized by Bardeen\cite{Bardeen} in 1980's. In the present two scalar 
model, however, if the $\phi^2\sigma^2$ term is radiatively induced 
with $O(1)$ coefficient, the (mass) hierarchy is immediately broken 
since $\VEV{\sigma}=M$ is of Planck or GUT energy scale.  
We have observed based on Ghilencea's explicit computation 
that the coupling constant $\lambda_m$ of the $\phi^2\sigma^2$ term 
remains of order $O(\varepsilon)$ at one-loop. The general reason for it is 
that the coupling between $\phi^2$ and $\sigma^2$ totally 
disappears in the 2-scalar model if $\lambda_m=\varepsilon\bar\lambda_m$ vanishes 
at tree level, so it must be proportional to $\varepsilon$ at any loop 
level\cite{ShapoZen}.  

If the gravity interaction is taken into account, we need additional 
reasoning, since $\phi^2$ and $\sigma^2$ can couple through gravity without 
factor $\lambda_m$. 
The point is that $\sigma^2$ can couple to the gravity loop directly 
via $\sqrt{-g}\sigma^2R$ whereas $\phi^2$ can couple to it only through the 
kinetic term $\sqrt{-g}g^{\mu\nu}\partial_\mu\phi\partial_\nu\phi$ unless we use the 
$O(\varepsilon)$ $\lambda_m$ interaction $\sqrt{-g}\lambda_m\phi^2\sigma^2$. 
If we use the kinetic term interaction vertex in which $\phi$ is accompanied 
by a derivative, the gravity loop graph will induce the term 
of the form $\sqrt{-g}\sigma^2\partial_\mu\phi\partial^\mu\phi$. It is of dimension 6 and should be 
divided by a mass dimension-2 quantity $M^2$. But we have no such a 
dimensionful parameter in this quantum scale invariant theory and 
only field that can appear in the denominator is the dilaton field $\sigma^2$, 
so it eventually gives just the Higgs kinetic term. 

Another worry is whether the gravity loop might induce the $\sigma^4$ term 
with $O(1)$ or even $O(\varepsilon)$ or not. The worry is only the gravity loops 
and only the term $\sqrt{-g}\sigma^2R$ is relevant. Then $\sigma^2$ plays the role of 
an overall factor in front of the action just like inverse of the 
Planck constant, $1/\hbar$. So the $L$-loop gravity graphs can produce 
only the term proportional 
to $(\sigma^2)^{1-L}$, and hence no positive power of $\sigma^2$ terms are 
induced by gravity loops.

\subsection{origin of the hierarchy}

In the above wishful scenario in Section 3.1, 
we have ``realized" {\em large gauge hierarchies} simply by 
assuming {\em tiny} parameters
\Blue{$\varepsilon_1\simeq 10^{-28}$ and $\varepsilon_2\simeq 10^{-34}$}: 
\begin{eqnarray}
V_1= \half\lambda_2\left( h^\dagger h - \varepsilon_1\Phi_1^2 \right)^2
\quad \hbox{and}  \quad 
V_2\supset\frac14 \lambda_2\left( \tr(\varphi^\dagger\varphi ) 
-\varepsilon_2\Phi_1^2 \right)^2\,.
\nonumber
\end{eqnarray}

However, the chiral symmetry breaking scale $\varepsilon_2\VEV{\Phi_1}$, 
for instance, can usually be explained by the running coupling 
as follows; if GUT is assumed, the SU(3) gauge coupling 
$\alpha_3=g^2_3/4\pi$ at scale $\sqrt{\varepsilon_0}M\equiv M_1$ 
evolves as $\alpha_3(\mu)$ a la RGE as the scale $\mu$ changes, and reaches 
to the $O(1)$ critical coupling $\alpha^{\rm cr}_3 \simeq 1$ at scale 
$\mu\simeq \Lambda_{\rm QCD}$ to break the 
chiral symmetry, so that $\sqrt{\varepsilon_2}M_1\simeq \Lambda_{\rm QCD}$. 
Thus the relation between the GUT scale $M_1$ and QCD scale 
$\Lambda_{\rm QCD}$ is fixed by the running gauge coupling constant $\alpha_3(M_1)$ 
at scale $M_1$ as
\begin{eqnarray}
\mu{d\over d\mu}\alpha_3(\mu)= 2b_3\,\alpha_3^2(\mu) \quad &\rightarrow&\quad 
{1\over\alpha_3(\mu)} ={1\over\alpha_3(M_1)}-b_3\ln{\mu^2\over M_1^2}  \nn
\quad &\rightarrow&\quad 
{1\over\alpha_3^{\rm cr}} ={1\over\alpha_3(M_1)}-b_3\ln{\Lambda^2_{\rm QCD}\over M_1^2} \,, \nonumber
\label{eq:epsilon2}
\end{eqnarray}
where $\alpha_3^{\rm cr}=O(1)$ quantity like $\pi/3$, so explains 
the huge hierarchy: 
\begin{equation}
\varepsilon_2={\Lambda^2_{\rm QCD}\over M_1^2}=\exp{1\over b_3}\Bigl(
{1\over\alpha_3(M_1)}-{1\over\alpha_3^{\rm cr}}\Bigr).
\end{equation}
This is the usual explanation. 

{\em In quantum SI theory}, 
$\alpha_3(M_1)$ here, probably, 
should be replaced by $M_1$-independent initial gauge coupling 
$\alpha_3^{\rm init}$, while the initial scale $M_1^2$ should be replaced 
by the dilaton field VEV $\VEV{\sigma}^2$. 
Then 
\begin{equation}
{1\over\alpha_3^{\rm cr}}-{1\over\alpha_3^{\rm init}}
= -b_3\ln {\Lambda^2_{\rm QCD}\over\VEV{\sigma}^2}
\end{equation}
so that the QCD scale $\Lambda_{\rm QCD}$ is always scaled with the dilaton VEV 
$\VEV{\sigma}$. 

This hierarchy should show up in the effective potential. 
Since $\Lambda^2_{{\rm QCD}}$ here should stand for the VEV $\varphi^\dagger\varphi$ of 
the chiral sigma model scalar field $\varphi$, we suspect that 
we should be able to derive an effective potential of quasi 
Coleman-Weinberg type like 
\begin{equation}
V(\sigma, \varphi)= {(\varphi^\dagger\varphi)^2\over64\pi^2}\left(-b_3\ln{\varphi^\dagger\varphi\over\sigma^2}
+{1\over\alpha_3^{\rm init}}-{1\over\alpha_3^{\rm cr}}\right)^2\,.
\end{equation}
Note that this form of SI potential is devised such that it has a 
non-vanishing field stationary point at 
$\VEV{\varphi}^\dagger\VEV{\varphi}/\VEV{\sigma}^2=x_0$ satisfying $b_3\ln x_0=
(\alpha_3^{\rm init})^{-1}-(\alpha_3^{\rm cr})^{-1}$.

\section{Conclusion}

I have shown in this talk that the scale invariance gives a 
natural mechanism for 
guaranteeing the vanishing vacuum energy at each step of hierarchical 
successive spontaneous symmetry breakings, at least in the classical 
field theory. 
I also explained the Englert-Truffin-Gastmans prescription which, I 
called quantum scale-invariant renormalization, preserves 
the scale-invariance also in quantum field theory. 
Even with such a prescription, however, 
the radiative corrections lift the flat directions of the 
potential and leave only the origin $\phi_i=0$ in field space 
as the stationary point, unless the superfine tuning of the 
coupling constants is made. Therefore, the scale-invariance alone 
is not sufficient for realizing the vanishing vacuum energy. 

I believe that the scale invariance is the right direction for solving the  
CC problem, but something is still missing.
We need yet another {\em symmetry} or a {\em mechanism} to realize
\begin{eqnarray}
\hbox{Spontaneous SI breaking} & =&  \hbox{Non-vanishing field VEV} \nn
 &=& \hbox{$\exists$ flat direction of $V(\phi)$}\,. \nonumber
\end{eqnarray}


\vspace{5ex}
\noindent
{\large\bf Acknowledgments}

\vspace{2ex}
I would like to thank Jisuke Kubo, Ichiro Oda and Dumitru Ghilencea 
for valuable discussions on scale invariant theories. I also owe to 
Makoto Kobayashi, Naoshi Sugiyama and Misao Sasaki for discussions 
on thermal history of the universe in the early stage of this work. 
I am also grateful to Toshihide Maskawa, Nobuyoshi Ohta, Ikuo Sogami, 
Shotaro Shiba, Naoki Yamatsu, Masato Yamanaka and other colleagues 
at Maskawa Institute, 
Kyoto Sangyo University for critical discussions and encouragement. 
This work is supported in part by Japan Society for the Promotion of 
Science (JSPS) Grant-in-Aid for Scientific Research (C) Grant Number 
JP18K03659.


\end{document}